%% file: main.tex
\documentclass[conference]{IEEEtran}
\IEEEoverridecommandlockouts
\usepackage[utf8]{inputenc}
\usepackage{lscape}
\usepackage{graphicx}
\usepackage{booktabs}
\usepackage{longtable}
\usepackage{threeparttable}
\usepackage{threeparttablex}

\usepackage[T1]{fontenc}

\newcounter{magicrownumbers}

\usepackage{nameref}
\usepackage{tikz}
\usepackage[bottom]{footmisc}

\input{packages}

\input{commands}

\begin{document}

\title{Go with the Flow? A Large-Scale Analysis of Health Care Delivery Networks in the United States Using Hodge Theory\\
\thanks{*Corresponding author.}
}

\author{\IEEEauthorblockN{Thomas Gebhart}
\IEEEauthorblockA{\textit{Computer Science and Engineering} \\
\textit{University of Minnesota}\\
Minneapolis, MN USA \\
gebhart@umn.edu}
\and
\IEEEauthorblockN{Xiaojun Fu}
\IEEEauthorblockA{\textit{School of Physics and Astronomy} \\
\textit{University of Minnesota}\\
Minneapolis, MN USA \\
fuxxx366@umn.edu}
\and
\IEEEauthorblockN{Russell J. Funk\IEEEauthorrefmark{1}}
\IEEEauthorblockA{\textit{Carlson School of Management} \\
\textit{University of Minnesota}\\
Minneapolis, MN USA \\
rfunk@umn.edu}
}

\maketitle

\begin{abstract}
Health care delivery is a collaborative process, requiring close coordination among networks of providers with specialized expertise. Yet in the United States, care is often spread across multiple disconnected providers (e.g., primary care physicians, specialists), leading to fragmented care delivery networks, and contributing to higher costs and lower quality. While this problem is well known, there are relatively few quantitative tools available for characterizing the dynamics of care delivery networks at scale, thereby inhibiting deeper understanding of care fragmentation and efforts to address it. In this, study, we conduct a large-scale analysis of care delivery networks across the United States using the discrete Hodge decomposition, an emerging method of topological data analysis. Using this technique, we decompose networks of patient flows among physicians into three orthogonal subspaces: gradient (acyclic flow), harmonic (global cyclic flow), and curl (local cyclic flow). We document substantial variation in the relative importance of each subspace, suggesting that there may be systematic differences in the organization of care delivery networks across health care markets. Moreover, we find that the relative importance of each subspace is predictive of local care cost and quality, with outcomes tending to be better with greater curl flow and worse with greater harmonic flow.
\end{abstract}

\begin{IEEEkeywords}
topological data analysis, Hodge theory, health care
\end{IEEEkeywords}

\section{Introduction}
Relative to comparable countries, the United States spends far more on health care, nearly 18\% of its Gross Domestic Product in 2016 \citep{hartman2018national}. Yet it has little to show for that spending, ranking near the bottom of Western, industrialized nations on many critical outcomes, ranging from infant mortality to longevity \citep{papanicolas2018health}. While the problems facing the United States health care system are complex and manifold, many suggest that the fragmented nature of care delivery is likely to play a decisive role \citep{lala2020manifesto, holmes2013heart}. Care fragmentation occurs when the delivery of services to patients is spread across multiple, disconnected providers (e.g., physicians) \citep{kreindler2012silos, frandsen2015care}. In settings with greater care fragmentation, communication and coordination among care team members is more difficult and therefore less robust. Consequently, care fragmentation contributes to higher and lower quality 
\citep{romano2015association, funk2018association, sutcliffe2004communication, mehrotra2011dropping}. 

In this study, we leverage recent advances in topological data analysis and the growing availability of “big data” on health care delivery to study care fragmentation at scale. Using administrative claims data from the Medicare population, we map the structure of care delivery networks across United States health care regions (2014-2017), wherein edges track flows of patients among local physicians. Subsequently, we apply the discrete Hodge decomposition to these networks \citep{schaub2020random, strang2020applications, jiang2011statistical}, which enables us to decompose the observed patient flows into their local cyclic (curl flow), global cyclic (harmonic flow), and acyclic (gradient flow) components. We then examine associations between these three different flow patterns and measures of local care quality and spending. 

While our work is not the first to study the large-scale properties of care delivery networks, existing literature has tended to adopt a static approach, focusing on network structure \citep{landon2012variation, landon2018patient, casalino2015physician, barnett2012physician, pollack2012physician}, and overlooking the directed nature of the underlying edges. Our study underscores the dynamic nature of care delivery networks and shows that the flow of patients through them encodes important information that is not visible using common structural measures. Moreover, our results suggest that Hodge theory is a particularly valuable tool for characterizing flows of patients in large-scale care delivery networks, one that, as we describe in greater detail below, likely corresponds well to clinically meaningful care pathways.

\section{Background and Related Work}

Contemporary interest in care delivery networks can be traced at least to the 1970s, when Shortell published a series of influential studies using the lens of social exchange theory \citep{shortell1973patterns, shortell1971physician, shortell1974determinants, blau2017exchange}. In brief, this work suggested that physicians’ referral patterns (i.e., to whom they refer and how much) were a function of perceived costs and social rewards. Conceptualizing physician referrals as a form of social exchange was significant, as it both laid the groundwork for future modeling using the tools of social network analysis and underscored the directed nature of relationships among health care providers (e.g., some exchanges are reciprocal, some are not). 

While this early work made significant conceptual advances, it was limited empirically; most research relied on field surveys, and was therefore typically constrained to the study of a single hospital, health system, or community, and at most several thousand physicians \citep{shortell1975patient, starfield2002variability, javalgi1993physicians, shortell1973patterns, shortell1974determinants, shortell1974referral, franks2000physicians}. In recent years, the growing availability of large-scale administrative data—primarily from electronic medical records and insurance claims—has allowed researchers to overcome these limitations, thereby leading to renewed interest in the study of care delivery networks \citep{dugoff2018scoping}. As with earlier work, much of the focus within this “second wave” of literature has been on understanding the causes and consequences of variation in the “local” (i.e., ego) networks of individual providers \citep{landon2018patient, barnett2012physician, tannenbaum2018surgeon}. The advent of “big data” has also, however, enabled the modeling of care delivery networks at larger, more “global” levels, with characterizations of the social structure among all providers within a particular organizational or geographical context becoming increasingly common \citep{landon2012variation, ghomrawi2018physician, hollingsworth2015differences, hollingsworth2015assessing, kim2019informal, funk2018association, everson2018repeated, an2018analysis, casalino2015physician}.

As this brief review shows, large-scale analyses have added considerable new insight. Nevertheless, in the process of “scaling up,” contemporary research has overlooked many of the rich, qualitative characteristics of provider relationships that animated earlier work. Chiefly among those, “second wave” literature on care delivery networks has devoted little attention to the directed nature of relationships among health care providers. Instead, contemporary work typically models edges as undirected, thereby implicitly treating the flow of information, influence, and patients among providers as reciprocal, and of equal weight in both directions. While part of this tendency reflects the early stage of the field (and the need to begin with simpler models), it is also attributable in part to the comparatively limited set of tools available for characterizing the global structure of directed networks \citep{malliaros2013clustering, foster2010edge}.  

In light of this overview, a gap is evident. As a result of data limitations, “first wave” literature on care delivery networks was typically small scale, thereby limiting the ability of researchers to study global network properties, and raising concerns about the generalizability of reported findings. Fueled by the growing availability of large-scale administrative data, “second wave” literature has mostly overcome these limitations. Yet existing tools, which are largely designed for the analysis of undirected networks, has led recent research to overlook a critical, qualitative feature of care delivery networks---their directionality---and, as such, has contributed to a growing void between foundational theory and contemporary measurement. In what follows, we suggest that Hodge theory provides a useful framework for addressing this gap.

\section{Data}
Our primary data are derived from Medicare claims. Medicare is the largest insurer in the United States and is the primary payer of medical bills for older adults. Medicare data are widely used by clinical and health services researchers, as they provide an exceptionally rich picture of health care delivery. 98 percent of individuals aged 65 and older are enrolled in Medicare, and the data are estimated to capture 99 percent of all deaths within that age group. Bills (or claims) submitted by providers (e.g., doctors) to Medicare for reimbursement include detailed information about patient diagnoses, services provided, dates and locations of service, and the billing providers. After processing these claims, Medicare makes them available for research via its website (for public use files) and the Research Data Assistance Center (for restricted use files).\footnote{\url{http://www.resdac.org}} To ensure privacy, patient identifiers are scrubbed or encrypted; however, physician identifiers (specifically the National Provider Identifier, or NPI) are retained, and therefore may be linked to external data sources.

To map the structure of care delivery networks in regions across the United States, we use large-scale, publicly available data on the flow of patients among providers in the Medicare program. These data are exceptionally rich, encompassing hundreds of millions of provider-provider relationships among millions of unique providers across all 50 states, ranging from the year 2014 to 2017. The data structure was designed by CareSet Labs, a health information technology consultancy, in conjunction with the Center for Medicare and Medicaid Services, the government organization that runs Medicare. 

To help better understand the potential determinants and consequences of different patient flow patterns, we also collected data on local care quality and spending (all files made available by the Dartmouth Institute for Health Policy and Clinical Practice, known as “TDI”). Basic data on providers (e.g., practice locations) were obtained from the National Plan and Provider Enumeration System (NPPES). 

\section{Methods}

\begin{figure*}[t]
\centering
\includegraphics[width=1.7\columnwidth]{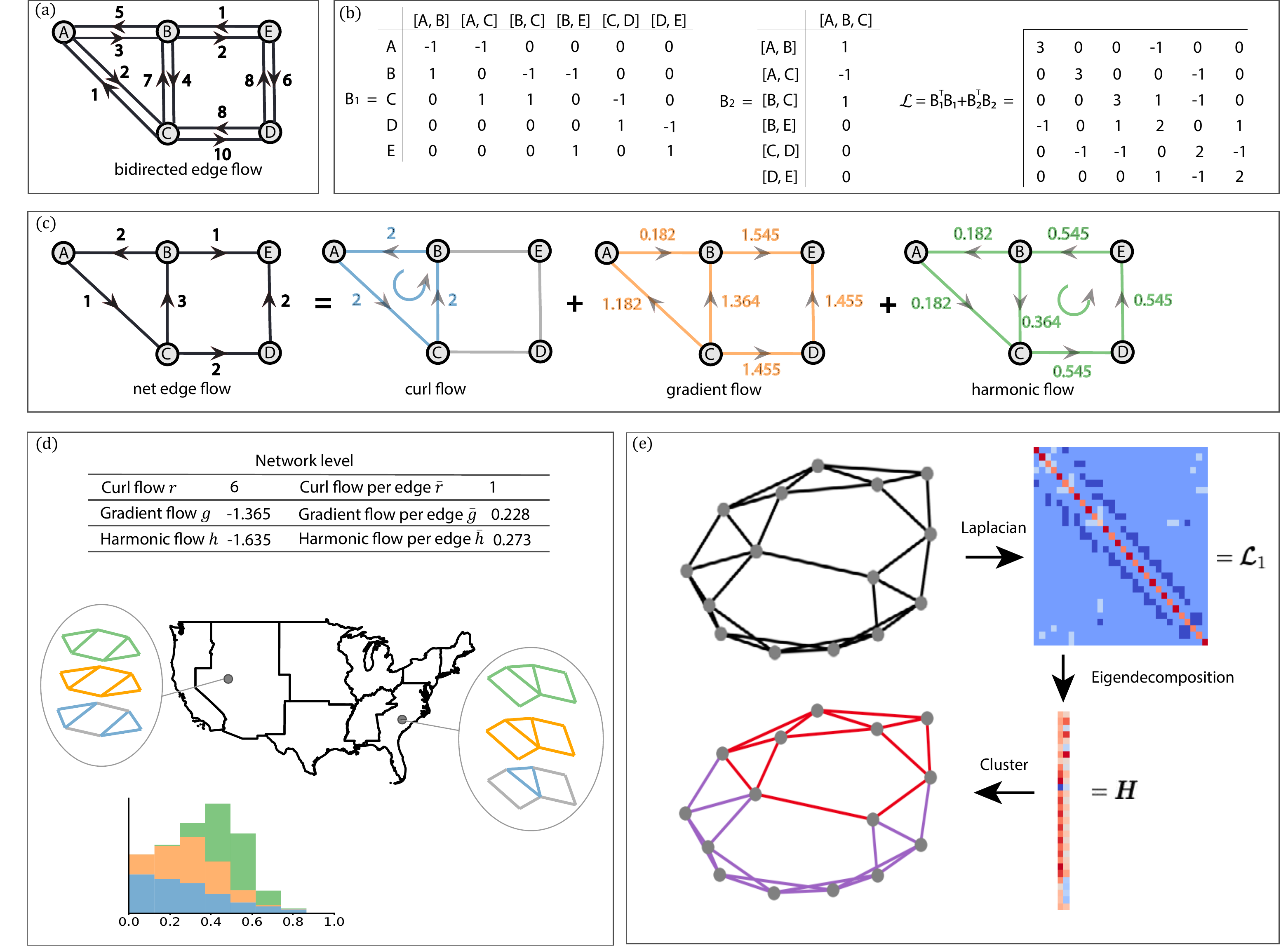}
\vspace{-0.05 in}
\caption{
\textsc{Methodological overview.} (a) Mapping of a care delivery network. Nodes correspond to providers. Directed edge flow from node A to node B corresponds to the number of referred patients from A to B within a defined time window. (b) The boundary operators $\mat{B}_1$ and $\mat{B}_2$. (c) Hodge decomposition on the net edge flow of the care delivery network shown in (a) into curl, gradient and harmonic flows. Curl flows are the net flows around 2-simplices, forming local cyclic flows. Gradient flows are acyclic and sum to zero around cycles. Harmonic flows sum to zero around 2-simplices but are non-zero along longer cycles, forming global cyclic flow. (d) To compare across time/space, we calculate the absolute values of sums of curl, gradient, and harmonic flows with all the edges at the network level, and normalize them by the network size. (e) Example of harmonic clustering using the Hodge $1$-Laplacian. We form the Laplacian $\lap_1$ from the network, compute its harmonic eigenvectors, then cluster these eigenvectors. The result corresponds to a clustering on the network edges.}
\label{fig:ill}
\vspace{-0.05 in}
\end{figure*}

\subsection{Mapping care delivery networks}\label{sec:data_methods}
The referral data (from CareSet Labs) are formatted as edge lists, one for each year of observation. Nodes correspond to providers (indicated by NPIs). Edges are recorded between pairs of providers when they bill for the same patients within a defined time window, and are weighted by the number of shared patients. For example, if NPI A saw 30 patients in one week, and 12 of those subsequently saw NPI B in the next week, we would record an edge between A and B with a weight of 12. As this example illustrates, there is a directionality to the edges, implied by the timing of patient visits, which motivates our view of these networks as tracking patient flows, and their characterization in the prior literature as “referral” networks \citep{an2018referral}.\footnote{For more details on how, precisely, the timing of patient visits is used to define directed edges, see \url{https://careset.com/docgraph-hop-teaming/}.}

The raw data include both individual (e.g., physicians) and organizational (e.g., hospital) providers. To facilitate interpretation and minimize undesirable heterogeneity, we subset our networks to include only individual providers. Moreover, we also limit our focus to only providers who are likely to be directly involved with patient care (e.g., we exclude radiologists). 

Health care delivery in the United States tends to be highly localized; patients are most likely to receive care from providers who are located close to where they live \citep{rosenthal2012geographic, weiss2020global, agha2019fragmented}. Therefore, consistent with prior work, we map care delivery networks within regions. We focus primarily on regions known as Hospital Service Areas (HSAs), which are widely used in health services research. Defined by the Dartmouth Institute, HSAs are geographic areas that correspond to discrete markets for hospital services.  For each observation year $\times$ HSA, we identify all local providers, based on practice addresses from the NPPES. We then map networks among those providers using the referral data. To complement these investigations, we also evaluate care delivery networks at larger geographic scales, including metropolitan areas and states. Finally, in several analyses, we consider the detailed geographic organization of care delivery networks, using precise latitude and longitude to position providers simultaneously in both physical and ``social'' space. Spatial coordinates were obtained by geocoding provider practice addresses using the Bing geocoding API. 

\subsection{Combinatorial Hodge Theory}

Classical Hodge theory, a celebrated theoretical bridge between differential geometry and topology, provides methods for characterizing the cohomology of smooth manifolds using tools from calculus and differential equations. Recent years have seen a resurgence of interest in applications of Hodge theory, stemming from the observation that many of the results within classical Hodge theory may be extended to discrete objects like networks, encoded as simplicial complexes. As we will observe below, Hodge theory applied to discrete data (often termed ``combinatorial Hodge theory'') leads to a natural generalization of more familiar network-theoretic notions of the graph Laplacian and spectral clustering, providing methods for characterizing higher-order processes taking place on the underlying networks.

Let $G = (\mathcal{V}, \mathcal{E})$ be a graph composed of vertex set $\mathcal{V}$ with cardinality $| \mathcal{V} | = n_0$ and edge set $\mathcal{E}$ with cardinality $| \mathcal{E} | = n_1$. The structure of the graph is often encoded linear-algebraically through the adjacency matrix $\mat{A}$ where elements $\mat{A}_{i,j} = 1$ if there exists an edge from node $i$ to node $j$ in the graph, $\mat{A}_{i,j} = -1$ if there exists an edge from $j$ to $i$ in the graph, and $\mat{A}_{i,j} = 0$ if $i$ and $j$ are not connected by an edge. When $G$ is undirected, each element of $\mat{A}$ is non-negative and $\mat{A}$ is symmetric. We can define the degree matrix of the graph as $\mat{D} = \diag{\mat{A}\mat{1}}$ such that $\mat{D}$ is a diagonal matrix of dimension $(n_0 \times n_0)$ with diagonal entries $\mat{D}_{i,i}$ representing the net number of edges leaving node $i$. With these two matrices, we define the graph Laplacian as $\mat{L} = \mat{D} - \mat{A}$. The structure of the graph Laplacian encodes many interesting properties about the underlying graph $G$, including the number of connected components of the graph, its sparsest cuts, and node centrality. The graph Laplacian may be normalized in a number of ways, providing additional insight into the structural properties of $G$. For example, the random walk normalization $\mat{L}^{\text{RW}} = \mat{I} - \mat{D}^{-1}\mat{A}$ represents the transition matrix of a random walker traversing the nodes of the graph with $(\mat{L}^{\text{RW}})^t$ encoding the occupancy probability of the walker after $t$ steps. The kernel of $\mat{L}$ is of particular interest. The dimension of $\ker\mat{L}$--the number of zero eigenvectors of $\mat{L}$--corresponds to the number of connected components of $G$. The associated eigenvectors, taking constant values on each component, are called the harmonic representatives. The celebrated Cheeger constant~\citep{cheeger1970lower} measures the work required to disconnect a graph as a bound on the smallest non-zero eigenvalue of $\mat{L}$. 

These notions of harmonicity and near-disconnectedness arising from the algebraic spectrum of the graph extend to higher-order network structures through the lens of Hodge theory. To see this, we require some machinery from algebraic topology. We will give a brief introduction to simplicial complexes and the Hodge decomposition in the following sections employing the notational conventions of \cite{schaub2020random}. For a gentler introduction to Hodge theory on graphs refer to \cite{lim2020hodge}, and for additional depth see \cite{horak2013spectra}.

\subsubsection{Simplicial Complexes}

Given a set of vertices $\mathcal{V}$, a $k$-simplex $\mathcal{S}^k$ is a subset of $\mathcal{V}$ of cardinality $k+1$. A simplicial complex $\mathcal{K}$ is a set of simplices with the property that if $\mathcal{S} \in \mathcal{K}$, then any subsets $\sigma \in \mathcal{S}$ is also in $\mathcal{K}$. A face of a simplex $\mathcal{S}^k$ is a subset of $S^k$ with cardinality $k$. If $S^{k-1}$ is a face of $\mathcal{S}^k$, we say $S^k$ is a co-face of $S^{k-1}$. From the preceding definitions, we see that a graph $G = (\mathcal{V}, \mathcal{E})$ may be interpreted as a simplicial complex with $0$-simplices $\mathcal{V}$ and $1$-simplices $\mathcal{E}$. Under this construction, $G$ is a relatively simple simplicial complex, lacking higher-order simplicial structure. One way to endow $G$ with more interesting topological structure is to transform it into a simplicial complex by computing its clique complex. This construction amounts to ``filling in'' any $k$-cliques in $G$ as $(k-1)$-dimensional simplices. This results in a simplicial complex with $1$-skeleton completely determined by the underlying graph, but simplices $\mathcal{S}^k$ for $k > 1$ determined by its clique structure.

Given an arbitrary but consistent orientation for each simplex, we may define the vector space $\mathcal{C}_k$ with coefficients in $\mathbb{R}$ and basis elements given by $\sigma_i^k$. Elements of $\mathcal{C}_k$ are called $k$-chains and are expressed as linear combinations of the basis elements $c_k = \sum_{i}\gamma_i\sigma_i^k$. Letting $n_k = | \mathcal{K}^k |$, we can represent each element of $\mathcal{C}_k$ by a vector $\vec{c} = (\gamma_1, \dots, \gamma_{n_k})^\top$. Endowing $\mathcal{C}_k$ with the standard $\ell_2$ inner product makes $\mathcal{C}_k$ a finite-dimensional Hilbert space. The space of $k$-chains $\mathcal{C}_k$ has a formal dual space of $k$-dimensional co-chains, denoted $\mathcal{C}^k$, composed of alternating functions $f: \mathcal{C}_k \to \mathbb{R}$. The spaces $\mathcal{C}_k$ and $\mathcal{C}^k$ are canonically isomorphic, so viewing these spaces from either viewpoint is acceptable. For our current applications, however, $\mathcal{C}^1$ has the intuitive interpretation as the space of edge flows on the graph $G$. In other words, $\vec{f} \in \mathbb{R}^{n_1}$ assigns a real-valued flow to each edge of $G$ with a negative value indicating flow opposite in direction across the orientation of the edge. 

We can define the linear boundary operator $\partial_k: \mathcal{C}_k \to \mathcal{C}_{k-1}$ as an alternating sum on the basis elements:
\begin{equation}
    \partial_k([i_0, \dots, i_k]) = \sum\limits_{j=0}^k (-1)^j[i_0, \dots, i_{j-1}, i_{j+1}, \dots, i_k]
\end{equation} which maps $k$-chains to a sum of its lower-adjacent boundary simplices. The image of this map $\im (\partial_k)$ defines the space of $(k-1)$-boundaries and its kernel $\ker (\partial_k)$ the space of $k$-cycles. This operator has a natural adjoint known as the coboundary map $\partial_k^\top: \mathcal{C}_k \to \mathcal{C}_{k+1}$, taking lower-adjacent simplices to their inclusion within $(k+1)$-simplices. Because these boundary (coboundary) maps are linear and applied to linearly-representable simplicial objects, we can encode them as matrices $\mat{B}_k$ ($\mat{B}_k^\top$). Of particular importance to this paper are the matrices $\mat{B}_1$ and $\mat{B}_2$. The boundary map $\mat{B}_1$ has rows indexed by $\mathcal{V}$ and columns indexed by $\mathcal{E}$ and is sometimes known in the graph-theoretic literature as the (signed) incidence matrix. The boundary map $\mat{B}_2$ has rows indexed by $\mathcal{E}$ and columns indexed by the $2$-simplices of $\mathcal{K}$. For $\mathcal{K}(G)$ constructed as the flag complex of $G$, $\mat{B}_2$ is indexed along the columns by the $3$-cliques of $G$, viewed as triangles in $\mathcal{K}^3$. An example of these boundary matrices can be found in Fig.\,\ref{fig:ill}(b).

\subsubsection{Hodge Laplacians and the Hodge Decomposition}

The boundary matrices $\mat{B}_k$ define an integer-indexed family of linear operators spanning the dimensions of $\mathcal{K}$. We define the $k$-dimensional Hodge Laplacian as 
\begin{equation}
    \lap_k = \mat{B}_k^\top \mat{B}_k + \mat{B}_{k+1}\mat{B}_{k+1}^\top. 
\end{equation} Of particular interest for our analyses is the Hodge 1-Laplacian: 
\begin{equation}\label{eq:1_laplacian}
    \lap_1 = \mat{B}_1^\top \mat{B}_1 + \mat{B}_2 \mat{B}_{2}^\top.
\end{equation} The 1-Laplacian represents a generalization of the standard graph Laplacian $\lap_0 = \mat{L} = \mat{B}_1 \mat{B}_1^\top$. An example of this matrix can be found in Fig.\,\ref{fig:ill}(b). Elements of the null space $\vec{h} \in \ker (\lap_k)$ are referred to as harmonic functions. These harmonics are intimately related to the topology of the simplicial complex over which $\lap_k$ is defined. The vector space $\mathcal{H}_k$ is composed of the elements in $\ker (\mat{B}_k)$ which are not in the image $\im (\mat{B}_{k+1})$, such that $\mathcal{H}_k = \mathcal{H}(\mathcal{K}, \mathbb{R}) = \ker(\mat{B}_k) / \im (\mat{B}_{k+1})$. The dimension of this space encodes the number of $k$-dimensional holes in the simplicial complex $\mathcal{K}$, also known as the $k$-dimensional Betti number $\beta_k$. Because $\lap_k$ is the sum of two positive semi-definite matrices, any $\vec{h} \in \ker (\lap_k)$ requires that both $\vec{h} \in \ker(\mat{B}_k)$ and $\vec{h} \in \ker(\mat{B}_{k+1}^\top)$. Therefore, the elements of $\ker (\lap_k)$ are representatives of the elements of $\mathcal{H}_k$ and, in coordination with standard linear-algebraic arguments\footnote{This follows from the direct sum decomposition of a vector space from its kernel as a linear map.}, we can decompose the vector space $\mathcal{C}_k$ as
\begin{equation}
    \mathcal{C}_k = \im(\mat{B}_{k+1}) \oplus \im (\mat{B}_k^\top) \oplus \ker (\lap_k)
\end{equation} which is known as the Hodge decomposition. On the space of edge flows $\mathcal{C}^1$, the decomposition 
\begin{equation}
    \mathcal{C}^1 \cong \mathcal{C}_1 = \im(\mat{B}_{2}) \oplus \im (\mat{B}_1^\top) \oplus \ker (\lap_1)
\end{equation} can provide additional insight into the data represented by the simplicial complex of the underlying graph. The subspace $\im (\mat{B}_2)$ is the curl subspace which consists of all weighted flows $\vec{r} \in \im (\mat{B}_2)$ that may be composed of local circulations along any $2$-simplex ($3$-clique). This is the discrete analog to the curl vector from calculus. The subspace $\im (\mat{B}_1^\top)$ is a weighted cut-space of edges consisting of linear-combinations of edges which disconnect the network. We may view these combinations as gradient flows $\vec{g} \in \im (\mat{B}_1^\top)$ which contains no cyclic component--the sum along any closed cycle is zero among elements of this subspace. The harmonic elements $\vec{h} \in \ker (\lap_1)$ are weighted global circulations that do not sum to zero around cycles but yet are inexpressible as linear combinations of curl flows around $2$-simplices. An example of this decomposition may be found in Fig.\,\ref{fig:ill}(c).

\subsubsection{Random-Walk Normalization}

Schaub et al.~\cite{schaub2020random} propose the random-walk normalized Hodge Laplacian as a generalization of the random-walk Laplacian on graphs, which itself encodes the transition matrix of a random walker on the underlying graph. In contrast to the graph Laplacian, the random-walk normalized form of the Hodge 1-Laplacian requires one to account for orientations of the simplices as the random walker traverses the simplicial complex. For a simplicial complex with boundary operators $\mat{B}_1$ and $\mat{B}_2$, the normalized Hodge 1-Laplacian is given by
\begin{equation}\label{eq:normalized_1_laplacian}
    \lap_1^{\text{RW}} = \mat{D}_2 \mat{B}_1^\top \mat{D}_1^{-1} \mat{B}_1 + \mat{B}_2 \mat{D}_3 \mat{B}_2^\top \mat{D}_2^{-1}
\end{equation} where $\mat{D}_2$ is the diagonal matrix of adjusted degrees of each edge ($\mat{D}_2)_{[i,j],[i,j]}$ = $\max(\text{deg}([i,j]),1)$, $\mat{D}_1 = 2*\text{diag}(|\mat{B}_1|\mat{D}_2\mat{1})$ is a diagonal matrix of weighted node degrees, and $\mat{D}_3 = \frac{1}{3}\mat{I}$ assigns degrees uniformly to $2$-simplices.

Comparing the unnormalized Hodge Laplacian in Equation~\ref{eq:normalized_1_laplacian} to the normalized form in Equation~\ref{eq:1_laplacian}, we see the effects of the network topology reflected in the added degree matrices in the normalization. The inverse $\mat{D}_2^{-1}$ term of the up-Laplacian reduces the influence of flows along edges which are incident to a large number of $2$-simplices. Similarly, the inverse node-degree matrix $\mat{D}_1^{-1}$ of the down-Laplacian reduces the influence of flows along edges which are incident to nodes of high degree before re-scaling by the $2$-simplex incidence once again. Decomposing edge flow $\vec{f} \in \mathbb{R}^{n_1}$ with respect to the standard inner product over $\lap_1^{\text{RW}}$ we see:
\begin{equation}
    \vec{f} = \im(\mat{D}_2^{-1/2} \mat{B}_2) \oplus \im(\mat{D}_2^{1/2} \mat{B}_1^\top) \oplus \ker(\mat{D}_2^{-1/2} \lap_1^{\text{RW}} \mat{D}_2^{1/2}).
\end{equation} Clearly, the decomposition of edge flow into the curl, gradient, and harmonic subspaces is affected by the incidence of edges among $2$-simplices. Therefore, we expect these effects to be most apparent in the decomposition of high-magnitude flows along edges which are incident to a large number of $2$-simplices. As a consequence, the difference in flow decomposition between the normalized and unnormalized flows to be especially prevalent in densely-connected networks whose edges take high flow values in the densest areas of the network.

This correlation between edge degree and flow magnitude is likely a feature of our network data; edges that move high referral volume are likely incident to nodes which are receiving or sending high referral volume among numerous other providers, therefore increasing the likelihood of these high-volume edges being incident to $2$-simplices\citep{granovetter1973strength}. Because of this, we opt to use the random-walk normalized Hodge Laplacian and its corresponding decomposition within our experiments. We will drop the ``RW'' superscript in the following sections, assuming all Hodge Laplacians are normalized, unless otherwise noted.  

\subsubsection{Harmonic Clustering}\label{sec:harmonic_clustering}

Many traditional graph-theoretic methods for characterizing the latent structure of networks like node embedding, spectral clustering, and community detection extend to simplicial complexes through the Hodge Laplacian~\citep{krishnagopal2021spectral, ebli2019notion, schaub2020random}. Of these methods, we expect clustering to be an especially insightful tool for simplifying the structure of care delivery networks. 

Ebli and Spreemann~\cite{ebli2019notion} introduce a useful generalization of spectral clustering to simplicial complexes which exploits the topological information embedded in the harmonic structure of the Hodge Laplacian which we now introduce. The eigendecomposition of $\lap_1$ may be written $\lap_1 = \mat{U}\mat{\Lambda}\mat{U}^\top$ where $\mat{\Lambda} = (\lambda_1, \lambda_2, \dots \lambda_{n_1})$ is a diagonal matrix of eigenvalues and $\mat{U} = (\vec{u}_1, \vec{u}_2 \dots, \vec{u}_{n_1})$ is a matrix with eigenvectors of $\lap_1$ as columns. Assume these eigenvectors and eigenvalues are aligned in decreasing order. The number of zero eigenvalues of $\lap_1$ correspond to $\beta_1$, the number of $1$-dimensional cavities of the underlying simplicial complex. The eigenvectors corresponding to these zero eigenvalues are the harmonic functions associated to $\lap_1$. We can collect these harmonic functions into a matrix $\mat{H} = (\vec{h}_1, \vec{h}_2, \dots, \vec{h_d})$ where $d$ corresponds to the degeneracy of the zero eigenvalues. We can then cluster $\mat{H}$ using any standard clustering method. However, as noted in \cite{ebli2019notion}, it is likely that the each $\vec{h}_i$ will lie along a low-dimensional subspace of $\mathbb{R}^d$. As such, subspace clustering methods are expected to produce the best results. We employ the subspace clustering approach proposed by ~\cite{you2016oracle} in our experiments.

Note that in clustering $\mat{H}$ directly, we are implicitly viewing the $1$-simplices $\vec{\sigma_i^1}$ as (unweighted) basis vectors formed as the matrix $\mat{\Sigma} = (\vec{\sigma}_1^1, \vec{\sigma}_2^1, \dots, \vec{\sigma}_{n_e}^1) = \mat{I}$. Although we do not explore such an approach in our experiments, it is possible to extend this clustering to account for the weighted flows on the simplicial complex. Such an approach shares significant conceptual overlap with the path embedding described in ~\cite{schaub2020random} which we leave as a potential avenue for future analyses of care delivery network data.

\subsection{Network-Level Measures}
Application of the Hodge decomposition to the care delivery network data will yield real-valued flows for each edge by subspace. For our study, however, we are also interested in more aggregate differences in the composition of patient flows across regions, at the network level. To enable such analyses, we define the following three network-level measures of flow, computed for each region $i$ and year $t$:

\begin{equation}\textrm{Gradient flow per edge } \bar{g}_{it} = \frac{1}{E}|\sum_{e}^{E} g_e|\end{equation}
\begin{equation}\textrm{Harmonic flow per edge } \bar{h}_{it} = \frac{1}{E}|\sum_{e}^{E} h_e|\end{equation}
\begin{equation}\textrm{Curl flow per edge } \bar{r}_{it} = \frac{1}{E}|\sum_{e}^{E} r_e|\end{equation}

\noindent where $E$ is the total number of edges in the network of region $i$ in year $t$ and $|\sum g_e|$, $|\sum h_e|$, and $|\sum r_e|$ indicate the absolute value of the sum of the gradient, harmonic, and curl flow at network level. We use the bar over $\bar{g}$, $\bar{h}$, and $\bar{r}$ to indicate the weighting by $E$. For an illustrative calculation, see Fig.\,\ref{fig:ill}(d).

\begin{figure*}[t]
\centering
\includegraphics[width=1.7\columnwidth]{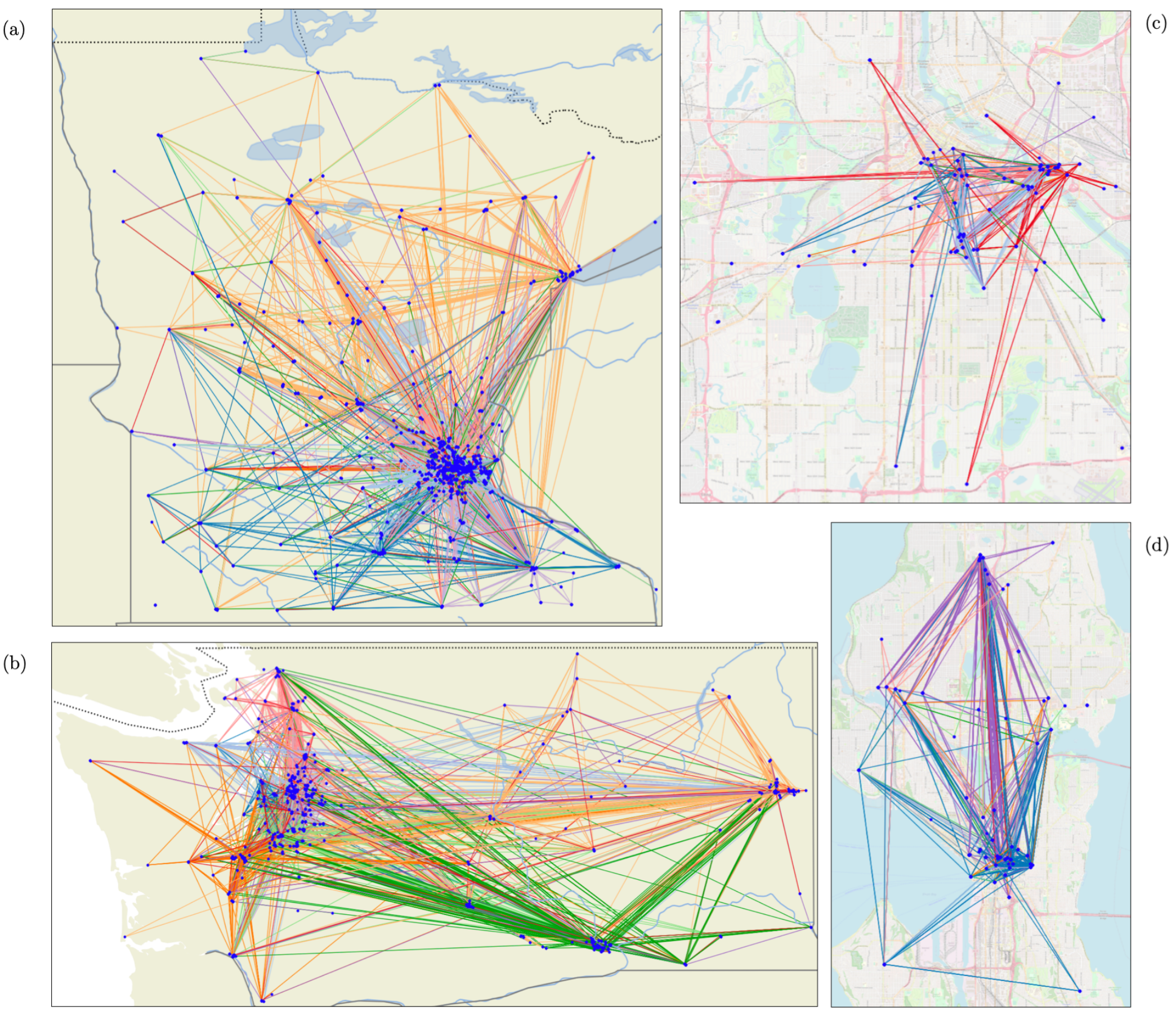}
\vspace{-0.05 in}
\caption{\textsc{Harmonic clustering of care delivery networks across two geographical scales.} Care delivery networks depicted for Minnesota (a), Washington (b), Minneapolis (c), and Seattle (d) as of 2017. Edges are colored by cluster assignment from the harmonic clustering method described in Section~\ref{sec:harmonic_clustering}}
\label{fig:harmonic_clustering}
\vspace{-0.05 in}
\end{figure*}

\section{Results}   

\subsection{Harmonic Clustering}

As an initial exploration of the topological structure of the care delivery network data, we choose a number of networks at different geographic scales to which we apply the harmonic clustering methods described in Section~\ref{sec:harmonic_clustering}. In order to visualize the geographic extent of these networks, we select providers whose address latitude and longitude could be successfully queried according to the methods described in Section~\ref{sec:data_methods}. For simplicity and ease of visualization, we fix the number of clusters to $10$ across all states and regions. The harmonic clustering assignments of the referral relationships for two states (Minnesota and Washington) and two major cities (Minneapolis and Seattle) are plotted with respect to their geographic structure in Fig.\,\ref{fig:harmonic_clustering}. This visualization of the care delivery networks illustrates the relationship between the cluster assignments and the geographic location of providers, especially at the state level. To reinforce this observation, we perform a simple K-means ($K=10$) clustering of the referral edges based only on the incident providers' latitudes and longitudes. We compute the adjusted mutual information (AMI) of the harmonic cluster assignments versus the K-means geographic cluster assignments for each of the four care delivery networks plotted in Fig.\,\ref{fig:harmonic_clustering}. At the state level, the AMI between the harmonic and geographic cluster assignments for Minnesota is $0.28$ and for Washington $0.43$. At the city level, the harmonic and geographic clustering AMI for Minneapolis is $0.45$ and for Seattle $0.16$. Clearly, the relationship between harmonic cluster assignment and location are real and non-random. However, the homological structure of these networks cannot be fully explained by provider proximity across referrals. 

Another possible driver of topological structure within care delivery networks is the ownership of individual providers within larger healthcare systems. Subsidiary providers within large healthcare systems are more likely to refer patients to other specialists within the system~\citep{patel2018closing}, therefore it is likely that the structure of referral relationships is driven in part by latent corporate cooperation. To investigate these system effects, we join provider health system ownership information from the publicly-available Agency for Healthcare Research and Quality's 2016 Compendium of U.S. Health Systems and assign clusters to network edges based on the unique system pairs which are adjacent across edges in the network. For example, given health systems A, B, and C, the possible clusters are (A,B), (B,C), (C,A), (A,A), (B,B), and (C,C). Computing the AMI between the harmonic clustering and these system cluster assignments, we find similar non-random relationships. For Minnesota and Washington, the AMI between the harmonic and system clusters are $0.25$ and $0.31$, respectively. For Minneapolis and Seattle, the AMI are $0.54$ and $0.30$, respectively.

While effects of geographic and system proximity on care delivery network topology are not surprising, it is nonetheless insightful to observe how these forces relate to the homological structure of care delivery networks. The harmonic clustering method described in Section~\ref{sec:harmonic_clustering} allows us to cluster referral relationships based on their incidence to homological structures within the larger network alone. This topology-aware clustering provides a general method for clustering referral relationships that does not depend on exogenous provider information like location or system ownership. We hypothesize that the homological structure which dictates these observed edge clusters, along with the harmonic flows generated by this structure, may vary by region and could relate to quality of care outcomes within the networks. We investigate these hypotheses in the following sections. 
\subsection{Variation in patient flows}
\begin{table*}[htbp]
\centering
\begin{threeparttable}
\vspace{-0.05 in}
\caption{
Characteristics of care delivery networks in representative regions in 2017
}
\begin{tabular}{ c  c  c  c  c } 
\toprule

Characteristics & Minneapolis, MN  & Albany, NY  & Joliet, IL & Portland, OR \\
\midrule
Total number of edges $E$ & 21995 & 19562 &  14369 & 19989  \\ 

Net flow $c$ & 18704.4 & 13845.8 & 7801.8 & 16064.2 \\

\midrule
Gradient flow $g$	 & 10933.5 & 7047.3 &  4251.6 &  9585.5  \\

Harmonic flow $h$	 & -9834.5 & -6813.1 &  -4171.5 &  -8670.0  \\

Curl flow $r$	 & 17605.4 & 13611.6 &  7721.7 &  15148.7  \\
\midrule
Gradient flow per edge $\bar{g}$  	 & 0.50 & 0.36 &  0.30 &  0.48  \\

Harmonic flow per edge $\bar{h}$  	 & 0.45 & 0.35 &  0.29 &  0.43  \\

Curl flow per edge $\bar{r}$	 & 0.80 & 0.70 &  0.54 &  0.76  \\
\bottomrule
\end{tabular}
\label{table:charac}
\vspace{-0.05 in}
\end{threeparttable}
\end{table*}

Using the approach outlined above (and illustrated in Fig.\,\ref{fig:ill}), we decompose patient flows in regional care delivery networks into their curl $r$, gradient $g$ and harmonic $h$ components, which we sum across edges to obtain a network-level measure, and normalize by network size (i.e., number of edges $E$). In Table\,\ref{table:charac}, we summarize the results for several representative regions, reporting total number of edges $E$, net flow $c$, gradient $g$, curl $r$, harmonic $h$, gradient per edge $\bar{g}$, curl per edge $\bar{r}$, and harmonic per edge $\bar{h}$ based on data from 2017. We observe notable differences in flow across regions. Minneapolis, MN and Portland, OR are fairly similar on all three dimensions of flow, and have more flow per edge on each subspace than Albany, NY and Joliet, IL, perhaps reflecting their status as larger metropolitan areas. Relative to Minneapolis and Portland, Albany and Joliet look relatively similar, although Albany has more flow per edge on all three dimensions, particularly in the curl subspace. 

\begin{figure}[htbp]
\centering
\includegraphics[width=1.0\columnwidth]{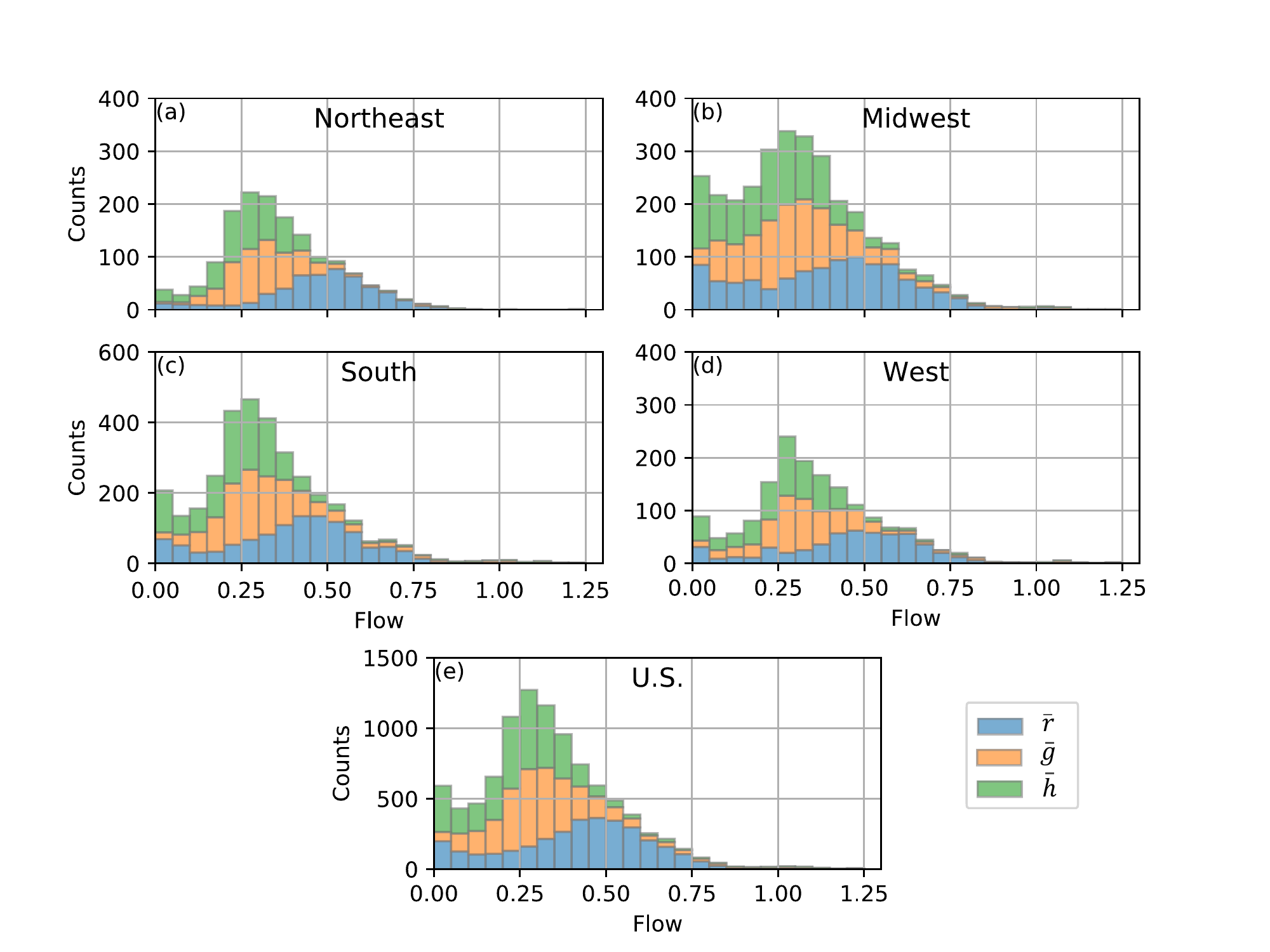}
\vspace{-0.25 in}
\caption{\textsc{Distribution of patient flows by subspace.} This figure shows histograms of the composition of patient flow across major regions of the United States: (a) Northeast, (b) Midwest, (c) South, (d) West, and (e) the overall United States. 
}
\label{fig:var}
\vspace{-0.05 in}
\end{figure}

To further examine variation in the composition of patient flow across regions, in Fig.\,\ref{fig:var}(a)-(d), we plot stacked histograms, showing gradient per edge $\bar{g}$, curl per edge $\bar{r}$, and harmonic per edge $\bar{h}$, separately for the Northeast, Midwest, South, West, and entire United States across four big regions in the United States. Values are based on flows computed at the HSA level, for the year 2017 (our most recent year of data). Comparing across different flow components, the mean of harmonic flow per edge $\bar{h}$ is the lowest in all regions, which naively, is expected, as the global cyclic characterized by the harmonic is relatively hard to form. Curl flow per edge $\bar{c}$, by contrast, assumes relatively larger values, which again may make sense in our context, as the formation of local cycles is probably easier and more natural in a care delivery network, where coordination among team members is important.

Perhaps more surprisingly, the distribution of $\bar{g}$, $\bar{r}$, and $\bar{h}$ varies across major regions of the United States.
For example, for curl flow, the Midwest has the lowest mean $0.39$ along with highest standard deviation $0.22$ compared with other regions, where the curl flow mean and standard deviation in the U.S. are $0.42$ and $0.21$ respectively. Interestingly, the Midwest also generates the lowest mean and the highest standard deviation in the harmonic flow. Among all the regions, the South has the most similar distribution pattern with overall U.S. These differences in the composition of patient flows may be indicative of different modes of health care delivery across the country.

\subsection{Association with spending and quality}

Up to this point, our results have shown that there is substantial variability in the composition of patient flows across regions. In this section, we consider whether this variation is correlated with health care spending and quality. To do so, we estimate a series of linear regression models. The unit of analysis in these models is the region $\times$ year. Each model includes three independent variables, which correspond to sum across of flow values assigned to each edge, separately for each of the three subspaces. To adjust for network size, we weight each sum by $E$ (i.e., the number of edges), after taking the absolute value. To adjust for temporal trends, each model includes year fixed effects. Finally, in supplemental models, we include as control variables measures of socioeconomic conditions, which have been shown in prior literature to be predictive of regional health care cost and quality \citep{zhang2020association, zhang2021social}, and are also related to the structure of care delivery networks \citep{landon2021assessment, ghomrawi2018physician, hollingsworth2015differences}. The specific measures included were median household income, unemployment rate, percent of the local population without a high school degree, and the relative proportion of residents who were black or Hispanic. Data for all socioeconomic measures were taken from the United States Census Bureau's American Community Survey. We use these measures as reference points against which to interpret the substantive significance of our flow measures.

Subsequently, we define several dependent variables, again using Dartmouth Atlas data. To capture spending, we estimate models of local reimbursements per beneficiary, with separate models for total, inpatient, and outpatient spending. \footnote{To account for regional differences in the cost of care, Medicare reimburses at different rates across regions. Our measure of spending is adjusted for these differences.} To capture quality, we estimate models of local readmission and ER visit rates among beneficiaries following discharge from the hospital after surgical treatment. All outcome measures are adjusted to account for regional differences in the age, sex, and race of the Medicare population. Prior health services research generally suggests that spending and quality are inversely related. Thus, negative coefficient estimates in models of spending and positive coefficient estimates in models of quality would be indicative of better performing regions.

{

\scriptsize

\setlength{\tabcolsep}{2pt}

\renewcommand{\arraystretch}{1}

\begin{table*}[htbp]\centering

\scalebox{1.0}{%

\begin{threeparttable}
\def\sym#1{\ifmmode^{#1}\else\(^{#1}\)\fi}
\vspace{-0.05 in}
\caption{Regressions predicting cost and quality}
\label{table:RegressionsWeighted}
\begin{tabular}{l*{5}{c}}
\toprule

 
                                                                                                                                       &\multicolumn{1}{c}{(1)}&\multicolumn{1}{c}{(2)}&\multicolumn{1}{c}{(3)}&\multicolumn{1}{c}{(4)}&\multicolumn{1}{c}{(5)}\\
                                             &\multicolumn{1}{c}{\shortstack{DV: Total \\ spending per \\ beneficiary \phantom{x} }}&\multicolumn{1}{c}{\shortstack{DV: Inpatient \\ spending per \\ beneficiary \phantom{x} }}&\multicolumn{1}{c}{\shortstack{DV: Outpatient \\ spending per \\ beneficiary \phantom{x} }}&\multicolumn{1}{c}{\shortstack{DV: Readmission \\ rate post-surgical \\ treatment }}&\multicolumn{1}{c}{\shortstack{DV: ER visit \\ rate post-surgical \\ treatment }}\\
\midrule
Gradient flow $\bar{g}$ per edge                   &          60.68   &          34.89   &        -288.66***&          -0.13   &          -0.37   \\
                                             &       (163.44)   &        (98.35)   &        (79.10)   &         (0.87)   &         (0.71)   \\
Harmonic flow $\bar{h}$ per edge                    &        1147.09***&        1137.64***&        2828.18***&           2.59** &           4.42***\\
                                             &       (321.92)   &       (201.14)   &       (160.58)   &         (1.31)   &         (1.11)   \\
Curl flow $\bar{r}$ per edge                       &       -1702.83***&       -1416.81***&       -2712.92***&          -2.80***&          -3.64***\\
                                             &       (216.07)   &       (135.18)   &       (105.29)   &         (0.52)   &         (0.51)   \\
Constant                                     &       10361.97***&        4726.87***&        2557.93***&          11.54***&          16.90***\\
                                             &        (57.38)   &        (37.14)   &        (29.81)   &         (0.16)   &         (0.16)   \\\midrule
Year fixed effects                           &            Yes   &            Yes   &            Yes   &            Yes   &            Yes   \\
\midrule
N                                            &          12952   &          12952   &          12952   &           6034   &           7776   \\
r2                                           &           0.08   &           0.05   &           0.22   &           0.01   &           0.03   \\

\bottomrule

\end{tabular}

\begin{tablenotes}

\item \emph{Notes:}  The level of analysis is the region $\times$ year. At the edge level, flow values are normalized using the approach described by \cite{schaub2020random}. After summing across edges to define the the region $\times$ year level measures, we divide each sum (i.e., $\bar{g}$, $\bar{h}$, $\bar{r}$) by the total number of edges in the network, as an adjustment for network size. Robust standard errors (clustered on region) are shown in parentheses; p-values correspond to two-tailed tests. 
\item {*}p<0.1; {**}p<0.05; {***}p<0.01

\end{tablenotes}
\vspace{-0.05 in}
\end{threeparttable}
}
\end{table*}

}%

The results of these analyses are shown in Table~\ref{table:RegressionsWeighted}. Recall that our flow variables are reported in units of average flow per edge. Beginning with Models 1-3, we find that both harmonic $\bar{h}$ and and curl $\bar{r}$ flow are significantly associated with all three measures of spending, but in opposite directions. When harmonic flow is greater, spending tends to be higher; when curl flow is greater, it tends to be lower. Based on estimates reported in Model 1, for an average region, a 1 SD decrease in harmonic $\bar{h}$ flow is associated with a decrease of \$186.83 in annual spending per beneficiary. While that may seem small, the aggregate savings by region (\$1,449,044) is more substantial. Turning to curl $\bar{r}$ flow, a 1 SD increase is associated with a decrease of \$354.75 in annual spending per beneficiary; for a region of average size, the corresponding savings works out to roughly \$2,751,425 per year. Relative to harmonic $\bar{h}$ and curl $\bar{r}$ flow, we find less consistent evidence of a relationship between gradient flow and spending (gradient flow is positively, though not significantly associated with total and inpatient spending, but negatively and significantly associated with outpatient spending). 

Models 4-5 show results of regressions predicting care quality. Consistent with our findings on spending, we find that greater curl $\bar{r}$ flow is associated with better outcomes, as indicated by statistically significant, negative coefficient estimates for readmission and ER visit rates following hospital discharge for surgical treatment. We also continue to find that greater harmonic flow is associated with worse outcomes, as indicated by positive and significant coefficient estimates for readmission and ER visit rates following surgical treatment. The associations we observe between flow and quality outcomes are also notable in substantive terms, although perhaps somewhat less remarkable than our findings on cost. For example, a 1 SD decrease in harmonic $\bar{h}$ flow is associated with a 4.46\% decrease in the predicted annual rate of ER visits following surgery.

In supplemental analyses, reported in Table~\ref{table:RegressionsWeightedCovariates}, we find substantively similar results when adjusting our models for local socioeconomic conditions. Inclusion of the three flow measures adds to the predictive power of all five regression models reported in the table, as indicated by statistically significant Wald tests (see the bottom of Table~\ref{table:RegressionsWeightedCovariates}). 

Models with socioeconomic variables included are useful as a point of reference when interpreting the substantive significance of coefficient estimates for our flow measures. For perspective, consider that a 1 SD decrease in the percent of the local population without a high school degree is associated with a decrease of \$465.76 in annual spending per beneficiary, which is roughly on par with the predicted savings associated with a similar decrease in harmonic flow reported above. Similarly, we find that the savings associated with a 1 SD decrease in the local unemployment rate is \$25.57 per beneficiary, which is substantially smaller in magnitude than the corresponding predictions for our flow measures.

Across models, the $R^2$ values are modest, ranging from 0.03 (predicting ER visit rates) to 0.22 (for outpatient spending per beneficiary), although they are comparable with prior work, and reflect the underlying difficulty of the prediction problem. When including the socioeconomic variables (Table~\ref{table:RegressionsWeightedCovariates}), model fit improves, though only slightly, with the new $R^2$ values ranging from from 0.06 (again predicting ER visit rates) to 0.29 (again for outpatient spending per beneficiary).

\section{Discussion}

Care fragmentation is a critical problem facing health care delivery in the United States. More than ever, patients receive medical attention not from a single physician, but rather from a host of different professionals, typically practicing across diverse organizational, institutional, and geographical settings, who often lack established collaborative relationships. While this model of care delivery may be beneficial in some ways (e.g., perhaps for innovation), it also poses significant challenges for care coordination. Recently, the growing availability of “big data” (e.g., insurance claims, electronic health records) has enabled unprecedented insight into health care delivery, thereby creating opportunities to better understand and ultimately address care fragmentation. 

Against this backdrop, we utilize a novel methodological framework from topological data analysis—the discrete Hodge decomposition—to study flows of patients among physicians in regional health care communities across the United States. Using this approach, we can characterize the nature of care delivery according to three distinctive patterns of flow: curl (local cyclic), harmonic (global cyclic), and gradient (noncyclic). Several interesting results emerge from our analyses. First, we find the topological structure which gives rise to the harmonic flows within the networks cannot be described entirely by geographic or healthcare system proximity. We find substantial variation in patterns of flow across regional health care communities. Moreover, we observe that this regional variation is not random, but rather seems to differ by broad regions of the United States (e.g., Northeast, West), perhaps corresponding to institutional differences in health care delivery. Second, we find that patient flow patterns contain useful information for partitioning care delivery networks into meaningful clusters. Finally, in regression analyses, we observe that regional patterns of patient flow are predictive of local health care cost and quality. Notably, across outcomes and model specifications, we find that while greater curl flow is associated with better local performance (i.e., lower cost, higher quality), greater harmonic flow is associated with worse performance. While these results are preliminary, these patterns seem plausible given the prior work from health services research. The movement of patients around global cycles (corresponding to “holes” or “gaps” in the network) seems likely to be problematic from a care coordination perspective, thereby potentially leading to higher cost, lower quality care. By contrast, the movement of patients around local cycles (as indicated by greater curl flow) seems likely to be conductive to close coordination among providers, allowing for better communication and more efficient care delivery. 

{

\scriptsize

\setlength{\tabcolsep}{2pt}

\renewcommand{\arraystretch}{1}

\begin{table*}[htbp]\centering

\scalebox{1.0}{%

\begin{threeparttable}
\def\sym#1{\ifmmode^{#1}\else\(^{#1}\)\fi}
\vspace{-0.05 in}
\caption{Regressions predicting health care cost and quality \\ with adjustment for local socioeconomic conditions}
\label{table:RegressionsWeightedCovariates}
\begin{tabular}{l*{5}{c}}
\toprule

 
                                                                                                                                                                                    &\multicolumn{1}{c}{(1)}&\multicolumn{1}{c}{(2)}&\multicolumn{1}{c}{(3)}&\multicolumn{1}{c}{(4)}&\multicolumn{1}{c}{(5)}\\
                                             &\multicolumn{1}{c}{\shortstack{DV: Total \\ spending per \\ beneficiary \phantom{x} }}&\multicolumn{1}{c}{\shortstack{DV: Inpatient \\ spending per \\ beneficiary \phantom{x} }}&\multicolumn{1}{c}{\shortstack{DV: Outpatient \\ spending per \\ beneficiary \phantom{x} }}&\multicolumn{1}{c}{\shortstack{DV: Readmission \\ rate post-surgical \\ treatment }}&\multicolumn{1}{c}{\shortstack{DV: ER visit \\ rate post-surgical \\ treatment }}\\
\midrule
Gradient flow $\bar{g}$ per edge                   &        -196.34   &        -127.45   &        -184.56** &           0.43   &          -0.09   \\
                                             &       (153.25)   &        (93.77)   &        (74.60)   &         (0.80)   &         (0.66)   \\
Harmonic flow $\bar{h}$ per edge                    &         557.08*  &         871.46***&        2748.45***&           0.67   &           2.42** \\
                                             &       (329.08)   &       (204.19)   &       (158.85)   &         (1.27)   &         (1.10)   \\
Curl flow $\bar{r}$ per edge                       &        -862.23***&       -1004.39***&       -2665.44***&          -1.51** &          -1.82***\\
                                             &       (238.69)   &       (146.99)   &       (111.30)   &         (0.60)   &         (0.59)   \\
Median household income (\$)                 &          -0.00   &           0.00   &          -0.00   &           0.00   &          -0.00***\\
                                             &         (0.00)   &         (0.00)   &         (0.00)   &         (0.00)   &         (0.00)   \\
Unemployment rate (\%)                       &           7.53   &           2.98   &         -50.70***&           0.08***&           0.12***\\
                                             &         (9.86)   &         (5.89)   &         (4.22)   &         (0.02)   &         (0.02)   \\
No high school degree (\%)                   &          71.01***&          44.43***&          -8.68***&           0.05***&           0.03***\\
                                             &         (5.19)   &         (3.00)   &         (1.53)   &         (0.01)   &         (0.01)   \\
Hispanic population (\%)                     &          -2.54***&          -1.25***&          -0.98***&          -0.00***&          -0.00   \\
                                             &         (0.68)   &         (0.37)   &         (0.21)   &         (0.00)   &         (0.00)   \\
Black population (\%)                        &           4.61***&           2.37***&           0.66***&           0.01***&          -0.00   \\
                                             &         (0.97)   &         (0.50)   &         (0.22)   &         (0.00)   &         (0.00)   \\
Constant                                     &        9227.76***&        4033.15***&        3007.61***&           9.94***&          15.35***\\
                                             &        (94.76)   &        (60.42)   &        (43.81)   &         (0.21)   &         (0.23)   \\\midrule
Year fixed effects                           &            Yes   &            Yes   &            Yes   &            Yes   &            Yes   \\
\midrule
N                                            &          12950   &          12950   &          12950   &           6034   &           7776   \\
R2                                           &           0.18   &           0.16   &           0.29   &           0.07   &           0.06   \\
\midrule Wald tests for flow predictors      &                  &                  &                  &                  &                  \\
F                                            &          11.08   &          19.79   &         208.88   &           3.82   &           4.14   \\
d.f.                                         &           3.00   &           3.00   &           3.00   &           3.00   &           3.00   \\
p-value                                      &           0.00   &           0.00   &           0.00   &           0.01   &           0.01   \\

\bottomrule

\end{tabular}

\begin{tablenotes}

\item \emph{Notes:}  The level of analysis is the region $\times$ year. At the edge level, flow values are normalized using the approach described by \cite{schaub2020random}. After summing across edges to define the the region $\times$ year level measures, we divide each sum (i.e., $\bar{g}$, $\bar{h}$, $\bar{r}$) by the total number of edges in the network, as an adjustment for network size. Robust standard errors (clustered on region) are shown in parentheses; p-values correspond to two-tailed tests. 
\item {*}p<0.1; {**}p<0.05; {***}p<0.01

\end{tablenotes}
\vspace{-0.05 in}
\end{threeparttable}
}
\end{table*}

}%

Several limitations should be kept in mind when interpreting our results. First, while prior work suggests that shared patients, as recorded in administrative claims data from Medicare, are predictive of self-reported professional ties among physicians, not all meaningful relationships will be captured in our data, and further, our data may also contain spurious relationships (i.e., records of ties in the form of shared patients where there is not established professional relationship). Second, our data are observational in nature, and the results of our regression analyses should therefore not be given a causal interpretation. For example, while we normalized our flow variables according to the number of edges in each network, additional differences in network size may be correlated with both our independent and dependent variables, thereby introducing residual confounding. Finally, while the associations we observe between regional flow patterns and cost and quality outcomes seem intuitively reasonable, more work needs to be done to validate the measures, ensuring that they correspond to meaningful differences in local care delivery.

Notwithstanding these limitations, we believe our study has implications for future research. Generally, our findings highlight the value of emerging methods in topological data analysis for the study of health care delivery. While our study is not the first to conduct large scale analyses of care delivery networks in the United States, prior work has generally used approaches from network science that are attentive to lower order structure (i.e., they characterize networks based on dyadic interactions among nodes). Although these approaches have led to valuable insights, the increasingly complex nature of modern health care delivery suggests that attention to higher order structure (i.e., characterizing networks based on larger groups of nodes) is likely necessary for developing a more complete understanding, a suggestion that is underscored by our observation of significant variability in the composition of patient flows across regions. 

\pagebreak

\bibliographystyle{plainnat}
\bibliography{bibliography}

\end{document}

%% file: packages.tex
\usepackage{booktabs} 
\usepackage{nicefrac}  
\usepackage{microtype}
\usepackage[utf8]{inputenc}

\usepackage{wrapfig}
\usepackage{multirow}

\usepackage{amssymb}
\usepackage{amsthm}
\usepackage{bbm}
\usepackage{bbm}
\usepackage{array}
\usepackage[colorlinks=false,urlcolor=purple,
linkcolor=purple,citecolor=purple]{hyperref}
\usepackage{amsmath}
\usepackage{mathtools}
\usepackage{float}

\usepackage{lscape}
\usepackage{graphicx}

\usepackage{booktabs}
\usepackage{longtable}
\usepackage{threeparttable}
\usepackage{threeparttablex}

\usepackage{mathrsfs}
\usepackage{graphicx}
\usepackage{color}
\usepackage[sort, numbers]{natbib}
\setcitestyle{square}
\usepackage{bm}

\usepackage{algorithm}
\usepackage{algpseudocode}
\usepackage{thmtools,thm-restate} 

\makeatletter
\def\paragraph{\@startsection{paragraph}{4}%
  \z@\z@{-\fontdimen2\font}%
  {\normalfont\bfseries}}
\makeatother

\usepackage{tikz}
\usetikzlibrary{decorations.markings, backgrounds, calc, positioning, shapes, shadows, arrows, fit, automata}
\usepackage{tikz-cd}
\usetikzlibrary{arrows.meta}
\tikzset{>={Latex}}

%% file: commands.tex
\newcounter{desccount}

\newcommand{\descref}[1]{\hyperref[#1]{#1}}

\newcommand{\im}{\operatorname{im}}

\theoremstyle{definition}

\makeatletter
\newcommand{\pushright}[1]{\ifmeasuring@#1\else\omit\hfill$\displaystyle#1$\fi\ignorespaces}
\newcommand{\pushleft}[1]{\ifmeasuring@#1\else\omit$\displaystyle#1$\hfill\fi\ignorespaces}
\makeatother

\renewcommand\vec{\bm}
\newcommand{\mat}[1]{{\bm{#1}}}
\newcommand{\diag}[1]{{\text{diag}\left({#1}\right)}}
\newcommand{\lap}{{\mat{\mathcal{L}}}}